Phase diagram studies for the growth of (Mg,Zr):SrGa$_{12}$O$_{19}$ crystals

Detlef Klimm[1,*], Bartosz Szczefanowicz[1,2,**], Nora Wolff[1,***], Matthias Bickermann[1]

[1]Leibniz-Institut für Kristallzüchtung, Max-Born-Str. 2, 12489 Berlin, Germany.

[2]Institute of Physics, Poznan University of Technology, Piotrowo 3, 60-965 Poznan, Poland

[*]corresponding author: detlef.klimm@ikz-berlin.de

[**] present address: INM – Leibniz Institute for New Materials, Campus D2 2, 66123 Saarbrücken, Germany

[***] present address: Helmholtz-Zentrum für Materialien und Energie, Hahn-Meitner-Platz 1, 14109 Berlin, Germany

Abstract
By differential thermal analysis a concentration field suitable for the growth of Zr, Mg co-doped strontium hexagallate crystals was observed that corresponds well with experimental results from Mateika & Laurien, J. Crystal Growth 52 (1981) 566–572. It was shown that the melting point of doped crystal is ca. 60 K higher than that of undoped crystals. This higher melting points indicates hexagallate phase stabilization by Zr, Mg co-doping, and increases the growth window, compared to undoped SrO–Ga$_2$O$_3$ melts.

Keywords: hexaferrite structure, thermal analysis, phase diagram, metastability, crystal growth

1. Introduction
Amongst the many pseudobinary compounds in the system SrO–Ga$_2$O$_3$, the composition of SrGa$_{12}$O$_{19}$ is closest to the component GaO$_{1.5}$ = ½ Ga$_2$O$_3$, with a molar fraction of GaO$_{1.5}$, $x$ = 0.9231 (Table 1). SrGa$_{12}$O$_{19}$ is isostructural to the mineral magnetoplumbite, (Pb,Mn$^{2+}$,Mg)(Fe$^{3+}$,Mn$^{3+}$)$_{12}$O$_{19}$, space group $P6_3/mmc$, which again belongs to the larger group of hexagonal ferrites, or "hexaferrites" [1]. Many of these materials possess strong and highly anisotropic persistent magnetic and electric moments, which makes them interesting as permanent magnets or even multiferroics. Crystal growth of Fe$^{3+}$ based hexaferrites is a challenge, because at the high melting points beyond 1500°C of these materials, partial reduction to Fe$^{2+}$ occurs; typically liquidus temperatures are reduced by foreign solvents like Na$_2$O to stabilize iron valency [2]. Resulting from the structural similarity, SrGa$_{12}$O$_{19}$ is a good substrate crystal for the epitaxial deposition of other hexaferrites [3]. Moreover, the chemical versatility of the magnetoplumbite structure allows doping of SrGa$_{12}$O$_{19}$ with luminescent ions such as Mn$^{2+}$ and Cr$^{3+}$ [4, 5].

The first publication of a phase diagram for the system SrO–Ga$_2$O$_3$ [6] showed that SrGa$_2$O$_4$ is the only intermediate compound with a congruent melting point. In more recent studies this system was redetermined and partially thermodynamically assessed [7, 8], with mainly similar results like the previous study [6] – but with the difference that the peritectic melting of SrGa$_{12}$O$_{19}$ was reported there at significantly lower temperature (Table 1). However, all studies agree with the observation that it melts peritectically under the formation of β-Ga$_2$O$_3$. A somewhat lower peritectic melting temperature 1540°C for SrGa$_{12}$O$_{19}$ and 1530°C for BaGa$_{12}$O$_{19}$ was reported elsewhere; both compounds form an isomorph solid solution series [9, 10]. For the SrO–Ga$_2$O$_3$ system, minor differences are reported mainly on the SrO side, the reader is referred to the PhD theses of Solak [11] and Richter [12].

As a result of peritectic melting, crystal growth of SrGa$_{12}$O$_{19}$ is only possible from melts with an excess of SrO, compared to the stoichiometry of the compound. According to the assessment of Zinkevich [8], this phase with a Ga$_2$O$_3$ molar fraction $x$ = 0.9231 is in equilibrium with the melt only between the peritectic points of SrGa$_{12}$O$_{19}$ ($x^{per}$ = 0.8002, $T^{per}$ = 1553°C) and the neighbouring phase SrGa$_4$O$_7$ ($x^{per'}$ = 0.7723, $T^{per'}$ = 1492°C). (The concentration data from [8] that are based on the components SrO and Ga$_2$O$_3$ were converted to SrO and GaO$_{1.5}$ which are used here.) With the lever rule, from these data a maximum yield $Y$ = (0.8002– 0.7723)/(0.9231– 0.7723) ≈ 18% for the growth of SrGa$_{12}$O$_{19}$ crystals from melts with excess SrO can be calculated.

First SrGa$_{12}$O$_{19}$ crystals with size up to 3 mm were grown by Haberey et al. [13] from fluxes with $x$ = 0.830, which means a slightly higher GaO$_{1.5}$ concentration than the peritectic point given by Zinkevich [8], $x^{per}$ = 0.8002. This difference indicates that supercooling of the melt might avoid the primary crystallization of β-Ga$_2$O$_3$. Later the same authors performed growth experiments with the addition of alkali molybdates or bismuth oxide. From melt solutions with Bi$_2$O$_3$ as component, SrGa$_{12}$O$_{19}$ crystals up to 15 mm diameter (useful area up to 30 mm$^2$) could be obtained. However, these crystals incorporated ca. 0.5 mol% Bi [14].

Significantly better and larger crystals were obtained by Mateika and Laurien [15]. They stated that the small concentration region in the pseudobinary system where SrGa$_{12}$O$_{19}$ crystallizes first (in their paper 0.7730 ≤ $x$(GaO$_{1.5}$) ≤ 0.8095, very similar to the data given above) can be increased, if Ga$^{3+}$ is substituted partially by small equimolar additions of Mg$^{2+}$ and Zr$^{4+}$. The partitioning coefficients of both ions was found to be k ≈ 1.05 >1, which suggests that the hexaferrite structure is stabilized. The possibility to substitute Ga$^{3+}$ by equimolar amounts of Mg$^{2+}$ and Zr$^{4+}$ was already earlier demonstrated for Gd$_3$Ga$_5$O$_{12}$ [16].

Table 1: Compounds in the pseudobinary system (1–x) SrO– x GaO$_{1.5}$. $T_f$ marks congruent or peritectic melting points, or peritectoid decomposition temperatures. $T_t$ are transition temperatures between different phases of one compound. For structural data of these compounds see e.g. Ropp [17].

| Formula | x | Remarks and reference |
| --- | --- | --- |
| SrO | 0.0000 | $T_f$ = 2665°C congruent [18] |
| Sr$_4$Ga$_2$O$_7$ | 0.3333 | $T_f$ = 1540°C [7], or 1476°C [6] peritectic |
| Sr$_7$Ga$_4$O$_{13}$ | 0.3636 | $T_f$ = 1490°C [7] peritectic, not found here |
| Sr$_{10}$Ga$_6$O$_{19}$ | 0.3750 | structure reported from [19, 20] |
| Sr$_3$Ga$_2$O$_6$ | 0.4000 | $T_f$ = 1230°C [7] peritectoid |
| Sr$_3$Ga$_4$O$_9$ | 0.5714 | $T_f$ = 1350°C [7], or 1322°C [6] peritectic |
| SrGa$_2$O$_4$ | 0.6667 | $T_t$ = 1430°C, $T_f$ = 1550°C [7], or 1580°C [6] congruent |
| SrGa$_4$O$_7$ | 0.8000 | $T_f$ = 1490°C [7], or 1422°C [6] peritectic |
| SrGa$_{12}$O$_{19}$ | 0.9231 | $T_f$ = 1550°C [7], 1553°C [8], or 1462°C [6] peritectic |
| Ga$_2$O$_3$ | 1.0000 | $T_f$ = 1800°C congruent [18] |

2. Experimental

Differential thermal analysis (DTA) with simultaneous thermogravimetry (TG) was performed using NETZSCH STA 449C "Jupiter" and STA 409CD thermal analysers. DTA/TG sample holders with Pt/Pt90Rh10 thermocouples and lidded platinum crucibles allowed measurements up to 1650°C in a flowing mixture of 20 ml/min Ar + 20 ml/min O$_2$. (Ga$_2$O$_3$ evaporates mainly under dissociation as Ga$_2$O, and SrO mainly as metallic Sr; and both reactions can be suppressed by adding O$_2$ to the atmosphere.) Usually the DTA samples were molten twice to ensure good mixing, and the second heating curves were used for further analysis. Unfortunately, under these experimental conditions the liquidus temperatures of mixtures close to the high melting components SrO and Ga$_2$O$_3$ (cf. Table 1) cannot be accessed, which prohibits good mixing and equilibration of DTA samples. Alternative DTA setups with higher maximum temperature cannot be used, because sample holder and/or furnaces contain then parts that are sensitive with respect to oxygen (e.g. from tungsten or graphite). Under such conditions, however, both components are prone to decomposition to metallic Sr or Ga, or Ga$_2$O suboxide, respectively, and subsequent evaporation. Ca. 50 different compositions spanning the whole range from pure SrO to pure Ga$_2$O$_3$ were prepared by melting together appropriate quantities of SrCO$_3$ and Ga$_2$O$_3$ powders (Alfa, 99.99% purity) in the DTA crucibles.

Prior to charging the samples into the DTA crucibles, the starting materials were checked for mass losses by emanating CO$_2$ (from carbonate calcination) or adsorbed volatiles such as traces of water. The samples themselves, with masses of at least 50 mg, were prepared on a balance with 0.01 mg resolution. This high accuracy ensures that concentration errors are insignificant.

In a second series MgO, ZrO$_2$, and an equimolar mixture of MgO + ZrO$_2$ was added to a (1 − $x$) SrO+$x$ GaO$_{1.5}$ mixture with $x$ = 0.857, that is close to the growth window of SrGa$_{12}$O$_{19}$. It was the aim of this series to reveal the influence of these dopants on the growth window.

## 3. Results and discussion

As mentioned in the previous section, the liquidus temperatures close to pure strontium or gallium oxide, respectively, are so high that evaporation from the sample prevents reliable thermal analysis. Not so in the centre of the system where a low eutectic (1326°C, $x$ = 0.49) between $Sr_{10}Ga_6O_{19}$ and $Sr_3Ga_4O_9$ results in low liquidus temperatures without significant evaporation (cf. Fig. 2). Nevertheless, another peculiarity made interpretation of DTA signals not straightforward there: It turned out that DTA curves were often not well reproducible, especially for compositions from the central region of the phase diagram. This is demonstrated for (1–$x$) SrO + $x$ GaO$_{1.5}$ mixtures with $x$ = 0.5549 (two subsequent heatings of one sample) and $x$ = 0.5855 (three heatings) in Fig. 1.

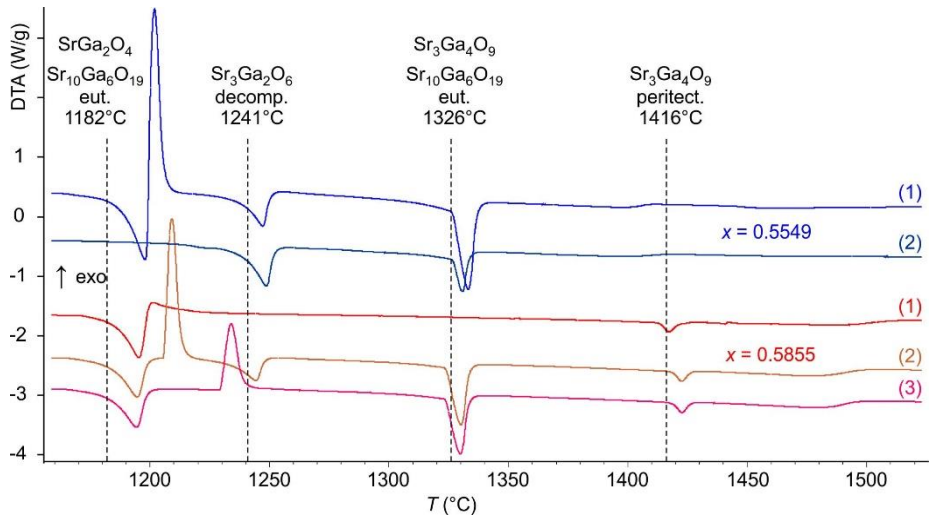

Figure 1: Subsequent DTA heating curves for identical samples with $x$ = 0.5549 (top two curves) or $x$ = 0.5855 (bottom three curves), respectively. Occasional exothermal peaks result from non-equilibrium that was obtained during previous cooling.

It is obvious that the curve (1) for sample $x$ = 0.5549, and curves (2) and (3) for sample $x$ = 0.5855, show exothermal peaks during these heating runs, which is untypical. All melting processes are endothermal events, but exothermal effects may occur if a sample is not in thermodynamic equilibrium and returns to equilibrium during heating.

For the $x$ = 0.5549 sample the peaks with onsets at 1241°C and 1326°C appear for both heating runs, because there equilibrium is obviously obtained, also for the upper curve. Not so the peak with onset at 1182°C which results from a non-equilibrium situation where $SrGa_2O_4$ and $Sr_{10}Ga_6O_{19}$ are coexisting. From Fig. 2 it can be seen that this is possible only if $Sr_3Ga_4O_9$ as well as $Sr_3Ga_2O_6$ are not formed. This can occur as a result of strong supercooling of both phases, which results in the non-equilibrium crystallization of their neighbours. Then, however, it is normal that these neighbour phases form together a lower eutectic (indicated by the dashed isotherm at 1182°C and the nonequilibrium prolongations of the liquidus lines in Fig. 2), than the equilibrium phases $Sr_3Ga_4O_9$ and $Sr_3Ga_2O_6$ would do.

Curve (1) for the $x$ = 0.5855 sample shows the same non-equilibrium eutectic, but immediately at the high temperature side of this peak a small exothermal bend occurs. No other effects appear until 1416°C, which is the peritectic melting temperature of $Sr_3Ga_4O_9$. This melting temperature was found here higher than reported in recent studies [6, 12], but we assume that these authors mixed up the eutectic at 1326°C with the peritectic melting of $Sr_3Ga_4O_9$. It should be noted that the composition of this sample is just 1.4% right from $Sr_3Ga_4O_9$, and hence this phase should be predominating there under equilibrium conditions. Only in curve (2) of this sample, $Sr_3Ga_2O_6$ is formed as a non-equilibrium phase first, which decomposes soon at 1241°C to $Sr_3Ga_4O_9$ and $Sr_{10}Ga_6O_{19}$, which then melt eutectically at 1326°C. The last heating curve (3) for this sample is similar to the previous one – with the difference that the exothermal jump into equilibrium occurs slightly later, and consequently the decomposition peak of $Sr_3Ga_2O_6$ cannot be observed.

After passing all DTA peaks, the *x* = 0.5855 curves show an upward bend near 1490°C. This indicates the liquidus temperature at this composition, because all melting processes are completed and the DTA curves return to their basis line. For the *x* = 0.5549 sample an analogous (but weaker) bend occurs near 1400°C because this composition is closer to the eutectic point.

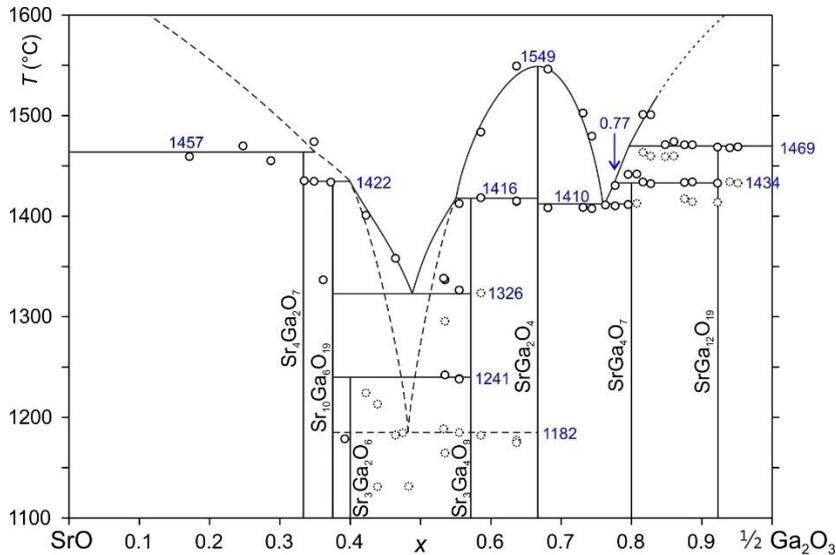

Figure 2: Experimental phase diagram Ga$_2$O$_3$–SrO with 7 intermediate compounds. Dashed liquidus lines indicate strong evaporation resulting from high temperatures close to the end members, or metastability near *x* = 0.5, respectively. Besides, one metastable eutectic at 1182°C is drawn.

A tentative phase diagram of the system ½ Ga2O3–SrO is shown in Fig. 2 which is partially based on the references [6, 7, 10, 11, 12], but complemented and corrected with experimental DTA points from this study. It is obvious that not all experimental points can be explained by the liquidus and isothermal lines in the diagram. However, additional non-equilibrium events can be expected to occur, e.g., if only one of the equilibrium eutectic phases at 1326°C is absent. Other effects, like the peaks on the 1410°C level right from *x* = 0.8, result from the initial crystallization of the hexagallate SrGa$_{12}$O$_{19}$. The remaining melt is depleted by Ga$_2$O$_3$ and its composition moves along the liquidus towards the eutectic point near *x* = 0.76, which produces then the corresponding peak also for compositions right from SrGa$_4$O$_7$.

In agreement with the Mateika & Laurien paper [15], Fig. 2 shows that the SrGa$_{12}$O$_{19}$ liquidus, and hence its crystallization window, is extremely narrow. Moreover, crystal growth is hampered there by the non-equilibrium crystallization of the neighbour phase SrGa$_4$O$_7$ [15]. As pointed out before, the occasional crystallization of non-equilibrium phases seems to be a general issue of the Ga$_2$O$_3$–SrO system.

In the magnetoplumbite crystal structure, the Ga$^{3+}$ ions reside in octahedral, bipyramidal, and tetragonal environments, and the five-fold coordinated Ga$^{3+}$ is randomly displaced from the centre of its trigonal bipyramidal coordination polyhedron along positive and negative directions of the *c*-axis [1, 21, 22]. Mateika & Laurien [15] succeeded to increase the growth window of SrGa$_{12}$O$_{19}$ by partial substitution of Ga$^{3+}$ (ionic radius $r[6]$ = 76; $r[4]$ = 61 pm [23]) by simultaneous substitution with identical amounts of Mg$^{2+}$ ($r[6]$ = 86; $r[4]$ = 71 pm) and Zr$^{4+}$ ions ($r[6]$ = 86; $r[4]$ = 73 pm), and crystals >1 cm$^3$ could be grown from a Sr$_{1.56}$Ga$_{10.40}$Mg$_{0.52}$Zr$_{0.52}$O$_{18.72}$ melt [15].

It was the purpose of further DTA measurements in this study to investigate how Mg$^{2+}$ and/or Zr$^{4+}$ doping influences relevant phase equilibria in the ½ Ga$_2$O$_3$–SrO system. From Fig. 2 it is evident that crystal growth of SrGa$_{12}$O$_{19}$ should be possible along its liquidus between the peritectic lines at 1469°C and 1434°C, which is a very narrow growth window. In three series of DTA measurements, to a SrO/Ga$_2$O$_3$ mixture with *x* = 0.8571 (where both peritectic peaks are strong) growing amounts of MgO only, ZrO$_2$ only, and of an equimolar MgO/ZrO$_2$ mixture were added.

Doping by exclusively MgO or ZrO$_2$ was not useful: In both cases the 1434°C peak (SrGa$_4$O$_7$ peritectic) is lowered by 10 K, but the 1469°C peak (SrGa$_{12}$O$_{19}$ peritectic) disappeared for additive levels around 4% – indicating instability of the hexagallate phase. Not so for equimolar MgO/ZrO$_2$ doping, which is shown in Fig. 3. It turns out that again the lower peritectic moves downwards, here by 20 K. Even more impressing is that the higher peritectic, which is the upper stability range of the hexagallate phase, shifts by >60 K upwards. As already pointed out by Mateika & Laurien [15], obviously the co-doping with Mg$^{2+}$/Zr$^{4+}$ increases the stability range. One can see from Fig. 3 that an upper useful co-doping level, is of the order $y = 0.1$, which means each 10% of MgO and ZrO$_2$ can be added. One can assume that the highly versatile coordinations [6], [5], [4] of Ga$^{3+}$ in the hexaferrite structure support the partial replacement of this ion by the Mg$^{2+}$/Zr$^{4+}$ dopant. Besides, the high number of four components leads at liquidus temperatures around 1500°C to a significant entropic stabilization of the (Mg,Zr):SrGa$_{12}$O$_{19}$ mixture phase.

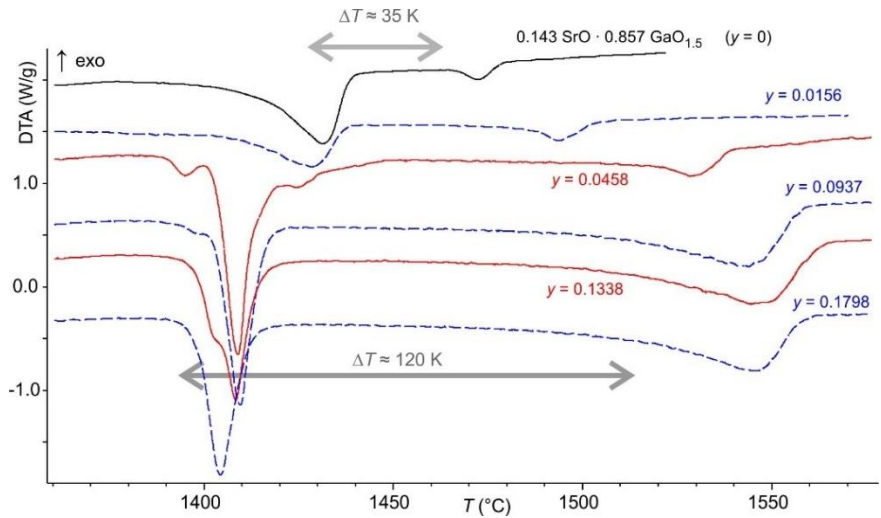

Figure 3: Starting from a (1–$x$) SrO+$x$ GaO$_{1.5}$ mixture with $x = 0.857$ (cf. Fig. 2), growing molar concentrations $y$ of a MgO:ZrO$_2$=1:1 mixture were added. This increases the difference between the lower SrGa$_4$O$_7$ and the higher SrGa$_{12}$O$_{19}$ peritectic decomposition significantly.

The graphical representation of this co-doping is not straightforward, because quaternary systems cannot be drawn without constrains in two dimensions. Mateika & Laurien [15] used a simplified concentration triangle with (MgO·ZrO$_2$)–Ga$_2$O$_3$–SrO as pseudocomponents or components, respectively. This is reasonable, because MgO and ZrO$_2$ are used only in the 1:1 molar ratio, and it is justified, because only the rim systems SrO–Ga$_2$O$_3$, SrO–ZrO$_2$, and MgO–Ga$_2$O$_3$ are relevant for the discussion. Fig. 4 a) is a similar presentation of this concentration triangle, with the difference that ½Ga$_2$O$_3$ and ½(MgO·ZrO$_2$) are defined as components. This has the benefit that all corners represent one single cation.

The further discussion may neglect the potential rim system MgO–SrO because this is simple eutectic without intermediate compounds, and hence no other phases that could crystallize first [24]. The other potential rim system ZrO$_2$–Ga$_2$O$_3$ is not known from the literature. However, simple Ga-Zr oxides do not exist and Ga–O–Zr bonds can be stabilized only with organic ligands [25]. Hence, one can assume that also the ZrO$_2$–Ga$_2$O$_3$ system is eutectic, like ZrO$_2$–Al$_2$O$_3$ [26]. Indeed, from Ga$_2$O$_3$ rich ternary melts with high MgO/ZrO$_2$ doping only MgGa$_2$O$_4$ crystallized in addition to SrGa$_{12}$O$_{19}$ and β-Ga$_2$O$_3$, and no Ga-Zr oxide phase was found [15]. Consequently, also the potential rim system ZrO$_2$–Ga$_2$O$_3$ can be neglected.

Both remaining rim systems that include MgO·ZrO$_2$ contain one intermediate compound with congruent melting behaviour: SrZrO$_3$ ($T_f$ = 2671°C, [27]) and MgGa$_2$O$_4$ ($T_f$ ≈ 1930…1950°C, [28, 29]). If intermediate compounds in ternary systems can coexist in equilibrium, tie lines can be drawn between them and the concentration triangle can be divided to partial systems. It is very common that such tie lines can be drawn between congruently melting phases, although exceptions are possible e.g. near ternary peritectic points [30]. In such cases, however, three solid phases should coexist, which was not reported in the literature [15, 31] so far.

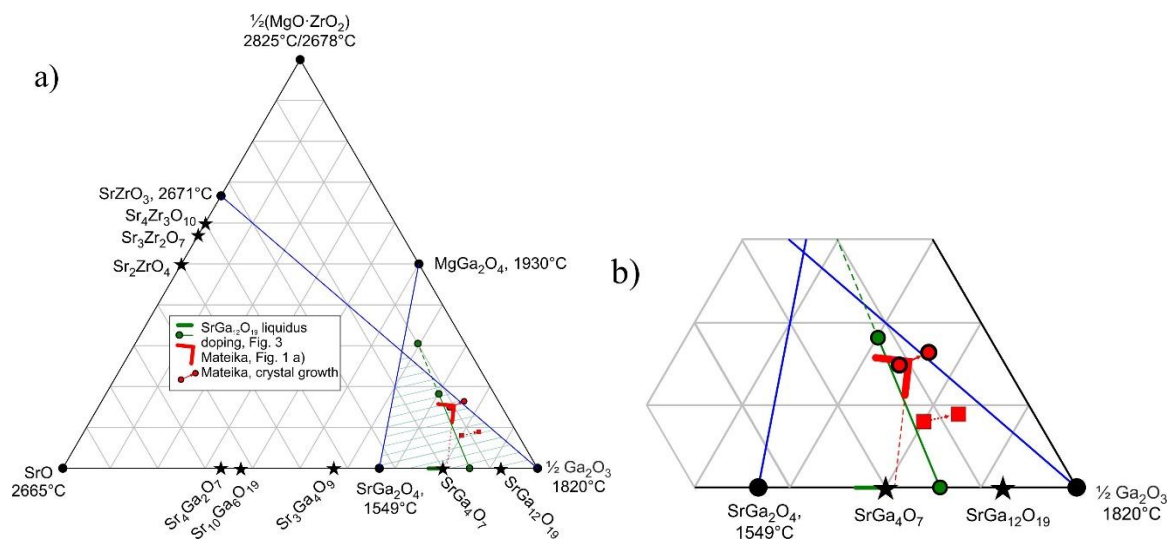

Figure 4: a) Concentration triangle SrO – ½ Ga$_2$O$_3$ – ½ (MgO·ZrO$_2$). Compounds with congruent melting points are marked by circles •, with peritectic melting by asterisks ⋆. Blue tie lines SrZrO$_3$–Ga$_2$O$_3$ and MgGa$_2$O$_4$–SrGa$_2$O$_4$ separate independent partial systems. Data of Mateika & Laurien [15] and from this study added. b) Ga$_2$O$_3$ rich corner of the triangle.

Fig. 4 a) shows the concentration triangle with these tie lines that separate independent partial systems. The considerations given above allow to conclude that for melts inside the shaded area, the whole crystallization path remains within this triangle, because this is a partial system. The triangle is enlarged in Fig. 4 b). The experiments by Mateika & Laurien resulted in the red corner as upper useful limits for the Ga$_2$O$_3$ and dopant concentrations (in reference [15] Fig. 1 a). In this publication the chemical composition of melts was compared with the composition of (Mg,Zr):SrGa$_{12}$O$_{19}$ crystals that were grown, and enrichment of the dopants in the crystal was found. Besides, the peritectic melting behaviour of SrGa$_{12}$O$_{19}$ requires melts with a smaller Ga$_2$O$_3$ concentration than the crystal. As a consequence, an upward right shift of the crystal compositions compared to the melts was observed [15].

Unfortunately, there is a contradiction: Fig 1 in [15] shows that (Mg,Zr):SrGa$_{12}$O$_{19}$ crystallizes only from melts inside the "red corner" in Fig. 4 of this article; but the melt composition Sr$_{1.56}$Ga$_{10.40}$Mg$_{0.52}$Zr$_{0.52}$O$_{18.72}$ that is given in Tab. 2 of the Mateika & Laurien paper corresponds to the left red square in Fig. 4, and the resulting crystal to the right square. We assume that concentration data were mixed up and can only guess that dopant concentration have to be doubled. Then the melt concentration lies almost exactly in the corner, and the result is (within the typical experimental error) almost exactly on the blue rim of the partial triangle. One can conclude that by trial and error Mateika & Laurien found a melt composition that is almost optimum for crystal growth in this system.

The DTA measurements that are shown in Fig. 3 are a confirmation: The growth window for (Mg,Zr):SrGa$_{12}$O$_{19}$ could be increased mainly by an increased stability of this hexagallate phase. This works well up to the $y$ = 0.0937 doping level. The starting composition of this doping series, $x$ = 0.857, and the useful upper doping level are marked by green circles in Fig. 4. Higher doping along the green dashed line is detrimental because the partial system is left.

4. Conclusions

Mateika & Laurien [15] identified a concentration field in the quaternary system SrO–Ga$_2$O$_3$–MgO–ZrO$_2$ were the growth of bulk (Mg,Zr):SrGa$_{12}$O$_{19}$ crystals is possible. With DTA measurements this concentration field was confirmed to be optimum, and a further optimization with respect to starting composition seems not possible. One technical error concerning concentration data in Tab. 2 of [15] was identified.

Acknowledgments

The authors thank Christo Guguschev for helpful discussions on this topic, and Steffen Ganschow for hints improving the manuscript. B.S. acknowledges support from the EU in the framework of the Erasmus+ program.